\documentclass[prb,a4paper,twocolumn,showpacs]{revtex4}
\input epsf
\usepackage{amssymb}
\usepackage{amsmath}
\begin{document}
\title{Crossover between ionic/covalent and pure ionic bonding in
magnesium oxyde clusters}
\author{F. Calvo}
\affiliation{Laboratoire de Physique Quantique, IRSAMC, Universit\'e Paul
Sabatier, 118 Route de Narbonne, F31062 Toulouse Cedex}
\begin{abstract}
An empirical potential with fluctuating charges is proposed for modelling
(MgO)$_n$ clusters in both the molecular (small $n$) and bulk ($n\to\infty$)
regimes. Vectorial polarization forces are explicitely taken
into account in the self-consistent determination of the charges.
Our model predicts cuboid cluster structures, in agreement with previous
experimental and theoretical results. The effective charge transferred
between magnesium and oxygen smoothly increases from 1 to 2, with an
estimated crossover size above 300 MgO molecules.
\end{abstract}
\pacs{%
36.40.Cg,
73.22.-f,
36.40.Ei
}
\maketitle

Crystalline magnesium oxyde is a purely ionic compound in which the Mg$^{Z+}$
and O$^{Z-}$ ions carry a charge $Z$ around $\pm 2$.\cite{causa} In gas phase,
the oxyde anion O$^{2-}$ is unstable and spontaneously decays into O$^-+$e
due to the strong electron-electron repulsion. As a result, the effective
atomic charge in the MgO molecule is much smaller than 2: independent {\em ab
initio} calculations by Ziemann and Castleman\cite{ziemann} and by Recio {\em
et al.}\cite{recio} found $Z\sim 0.8$. In the intermediate sizes regime,
(MgO)$_n$ clusters are thus expected to show intriguing properties due to a
partially covalent character of the chemical bonding. Beyond condensed matter
or molecular physics, these clusters received some special attention in the
astrophysics community, where they have been involved in the nucleation process
of dust in circumstellar shells around M-stars. 

Despite the vast amount of experimental\cite{ziemann,saunders,kim} and
theoretical\cite{ziemann,recio,veliah,li,aguado,coudray,kohler,roberts,wilson}
investigations on neutral or charged clusters, the way and the rate at which
chemical bonding evolves from ionic/covalent at small sizes toward purely ionic
in the bulk remains essentially unexplored. Because the electrostatic field
created by the ions does not vanish in finite systems, the highly polarizable
oxyde anion has a rather floppy and deformable outer electron cloud, which
could be responsible for a partial screening of the repulsion between cations.
However, the situation is complicated by the possible coordination-dependence
of the charge transferred locally.

Theoretical studies of (MgO)$_n$ clusters can be effectively separated into two
groups. {\em Ab initio} or density-functional theory (DFT) based calculations
have been performed on specific geometries, in a rather limited size
range.\cite{ziemann,recio,veliah,li,aguado,coudray} These works predict that
small clusters exhibit cuboid-like shapes similar to NaCl rocksalt
clusters. The apparent charge transferred, as estimated from Mulliken
populations, is indeed size- and coordination-dependent, and lies between
1 and 1.5 for $2\leq n\leq 13$.\cite{recio,coudray} More empirical methods have
also been used to predict optimal structures.\cite{ziemann,kohler,roberts}
Ziemann and Castleman\cite{ziemann} and, more recently, Roberts and
Johnston,\cite{roberts} have used the rigid ion model (Born-Mayer + Coulomb
interactions) potential with two possible values of the charge transferred.
When a charge $Z=1$ is taken, cuboids are preferentially found as the most
stable geometries. For $Z=2$, as in the bulk, small clusters show instead
hollow, fullerene-like structures. The effects of polarization have been
studied by K\" ohler and coworkers\cite{kohler} using the polarizable ion
model due to Rittner.\cite{rittner} Wilson\cite{wilson} investigated MgO
``nanotubes'' made of hexagonal (MgO)$_3$ rings stacked. For this he
developed a more sophisticated compressible-ion model\cite{cim} with
explicit coordination-dependent polarizabilities.

None of these empirical potentials account for the different charges
transferred in MgO clusters. Only in Ref.~\onlinecite{kohler} the authors
explicitely employed a size-dependent value of the charge $Z$, using an
arbitrary law $Z(n)=(2\zeta n+1)/(\zeta n+1)$. $\zeta$ was
taken such that the crossover $n^*$ between ionic/covalent and
purely ionic, for which $Z$ equals 1.5, occurs approximately at $n^*=20$. 
However, the low energy structures found by K\" ohler and coworkers
significantly deviate from regular cuboids or stacked hexagons in this size
range.\cite{kohler} An improved treatment of electrostatics and charge
transfer is provided by fluctuating charges (fluc-$q$)
potentials\cite{sawada,rappe,rick} based on the principle of
electronegativity equalization.\cite{sanderson} Such potentials have been
used in simulations of water\cite{rick,rick2,stern} and molten
salts,\cite{ribeiro} and have recently
proven valuable in describing the heterogeneous bonding in coated fullerenes
and nanotubes.\cite{wetting,sdickson} They have been extended to include
dipolar terms and the corresponding polarization.\cite{stern} In MgO clusters
atomic polarization cannot be neglected, and we provide here a self-consistent
treatment of these effects. Briefly, the system is made of $N$ magnesium
cations and $M$ oxygen anions, each carrying a charge $q_i$ and located at
the position vector ${\bf r}_i$. The total potential energy $V$ of the system
is written as $V=V_{\rm rep}+V_Q$. The repulsion interaction $V_{\rm rep}$ is
taken in the usual Born-Mayer format as a function of the distance $r_{ij}$
between ions $i$ and $j$:
\begin{equation}
V_{\rm rep}(\{ {\bf r}_i\}) = \sum_{i<j} D \exp(-\beta r_{ij}),
\label{eq:vrep}
\end{equation}
The total electrostatic energy $V_Q$ is expressed as
\begin{eqnarray}
V_Q(\{ {\bf r}_i\}) &=& \sum_i \left[\varepsilon_i q_i + \frac{1}{2}U_{ii}^0
q_i^2 - \frac{1}{2}\alpha_i {\bf E}_i^2\right] \nonumber \\
&& + \sum_{i<j} J_{ij}(r_{ij}) q_i q_j + \lambda\left( Q - \sum_i q_i\right).
\label{eq:vq}
\end{eqnarray}
$\varepsilon_i = \varepsilon_{\rm Mg}$ or $\varepsilon_{\rm O}$ are the
electronegativities associated with each atomic element. $J_{ij}$ is the
Coulomb integral between ions $i$ and $j$, taken in the Ohno representation
as\cite{ohno}
\begin{equation}
J_{ij}(r) = \left[ r^2 + (U_{ij}^0)^{-2}\exp(-\gamma_{ij} r^2)\right]^{-1/2}.
\label{eq:jij}
\end{equation}
$\alpha_i = \alpha_{\rm Mg}$ or $\alpha_{\rm O}$ are the atomic
polarizabilities, and ${\bf E}_i$ is the electric field vector felt by ion
$i$:
\begin{equation}
{\bf E}_i = \sum_{j\neq i} - q_j \frac{\partial J_{ij}}{\partial {\bf r}_{ij}}.
\label{eq:field}
\end{equation}
Finally, the last term in Eq.~(\ref{eq:vq}) includes a Lagrange multiplier
$\lambda$, which accounts for the conservation of the total charge $Q$ of the
system. Given an instantaneous set of positions $\{ {\bf r}_i \}$, the charges
$\{ q_i\}$ are found by minimizing Eq.~(\ref{eq:vq}) above. Thanks to the
linear dependence of the electric fields ${\bf E}_i$'s on the charges, the
expression of $V_Q$ is quadratic in the $q_i$'s, and its minimization can be
done readily using linear algebra. The charges are solution of the matrix
equation ${\bf C} {\bf X} = {\bf D}$, where ${\bf X}=\{ q_i, \lambda \}$ is
a $N+M+1$ vector and ${\bf D}$ has components $D_i = -\varepsilon_i$ for $i\leq
N+M$, and $D_{N+M+1}=Q$. The element $(i,j)$ of matrix ${\bf C}$ is expressed
as
\begin{equation}
C_{ij} = J_{ij} - \sum_k \alpha_k \left( \frac{\partial J_{ki}}{\partial {\bf
r}_{ki}}\cdot \frac{\partial J_{kj}}{\partial {\bf r}_{kj}}\right),
\label{eq:cij}
\end{equation}
with $J_{ii}=U_{ii}^0$. The polarizability of the cluster can be calculated
by imposing external electric fields and by computing the variations of the
electric dipole. The previous expressions need to be modified accordingly to
incorporate this effect.
This model has 11 independent parameters, including $D$, $\beta$, the
$\alpha_i$'s, the $U_{ij}^0$'s and $\gamma_{ij}$'s. Only the difference in
electronegativities $\Delta\varepsilon=
\varepsilon_{\rm Mg}-\varepsilon_{\rm O}$ is physically
relevant.\cite{rick} In order to make the model transferrable from the
molecular range up to the bulk, some constraints must be imposed on these
parameters. In the MgO diatomics, the equilibrium distance, charge transferred
and electric dipole are known (see for instance Ref.~\onlinecite{kohler} and
references therein). For this molecule, one must minimize the energy function
$V$ with respect to the Mg--O distance $r$. After some calculation we find
$V^{\rm MgO}(r) = De^{-\beta r} + V_Q^{\rm MgO}(r)$ with the electrostatic
term:
\begin{eqnarray}
&&V_Q^{\rm MgO}(r)= \nonumber \\
&&\frac{1}{2}\frac{(\Delta \varepsilon)^2}{(\alpha_{\rm Mg} +
\alpha_{\rm O}) [J'_{\rm MgO}(r)]^2 + 2J_{\rm MgO}(r) - U_{\rm MgMg}^0
- U_{\rm OO}^0},
\label{eq:vq2}
\end{eqnarray}
were we have employed the notation $J'_{\rm MgO}=dJ_{\rm MgO}/dr$.

In the (B1) crystal, the charge transferred $Z$ and lattice constant $a$ are
chosen as reference data. Using the assumption $J_{\rm MgO}(a)\sim 1/a$,
one finds the binding energy per ion:
\begin{eqnarray}
\frac{V^{\rm bulk}(a)}{N} &=& 6De^{-\beta a} + \frac{V_Q^{\rm bulk}(a)}{N},
\nonumber \\
\frac{V_Q^{\rm bulk}(a)}{N} &=& \frac{(\Delta \varepsilon)^2}{M/a - 
U_{\rm MgMg}^0 - U_{\rm OO}^0},
\label{eq:vbulk}
\end{eqnarray}
with $M$ the Madeling constant. For a set of parameters, the total energies
corresponding to the diatomics and to crystal must be minimized with
respect to $r$ or $a$, respectively. This is done numerically using the
Ohno form of the Coulomb integral, Eq.~(\ref{eq:jij}) above. The full
parameterization of the model can then be achieved by minimization of an
error function $\chi^2$, to reproduce some dimer and crystal properties. The
following values have been adopted: $D=6056$ eV, $\beta=4.89$~\AA$^{-1}$,
$U_{\rm MgMg}^0=0.46$, $U_{\rm OO}^0=1.13$, $U_{\rm MgO}^0=0.85$,
$\gamma_{\rm MgMg}=0.35$~\AA$^{-2}$, $\gamma_{\rm OO}=0.49$~\AA$^{-2}$,
$\gamma_{\rm MgO}=0.36$~\AA$^{-2}$, $\Delta \varepsilon=0.935$, $\alpha_{\rm
Mg}=0.18$~\AA$^3$, and $\alpha_{\rm O}=4.65$~\AA$^3$.

\begin{figure}[htb]
\setlength{\epsfxsize}{8.6cm}
\leavevmode \epsffile{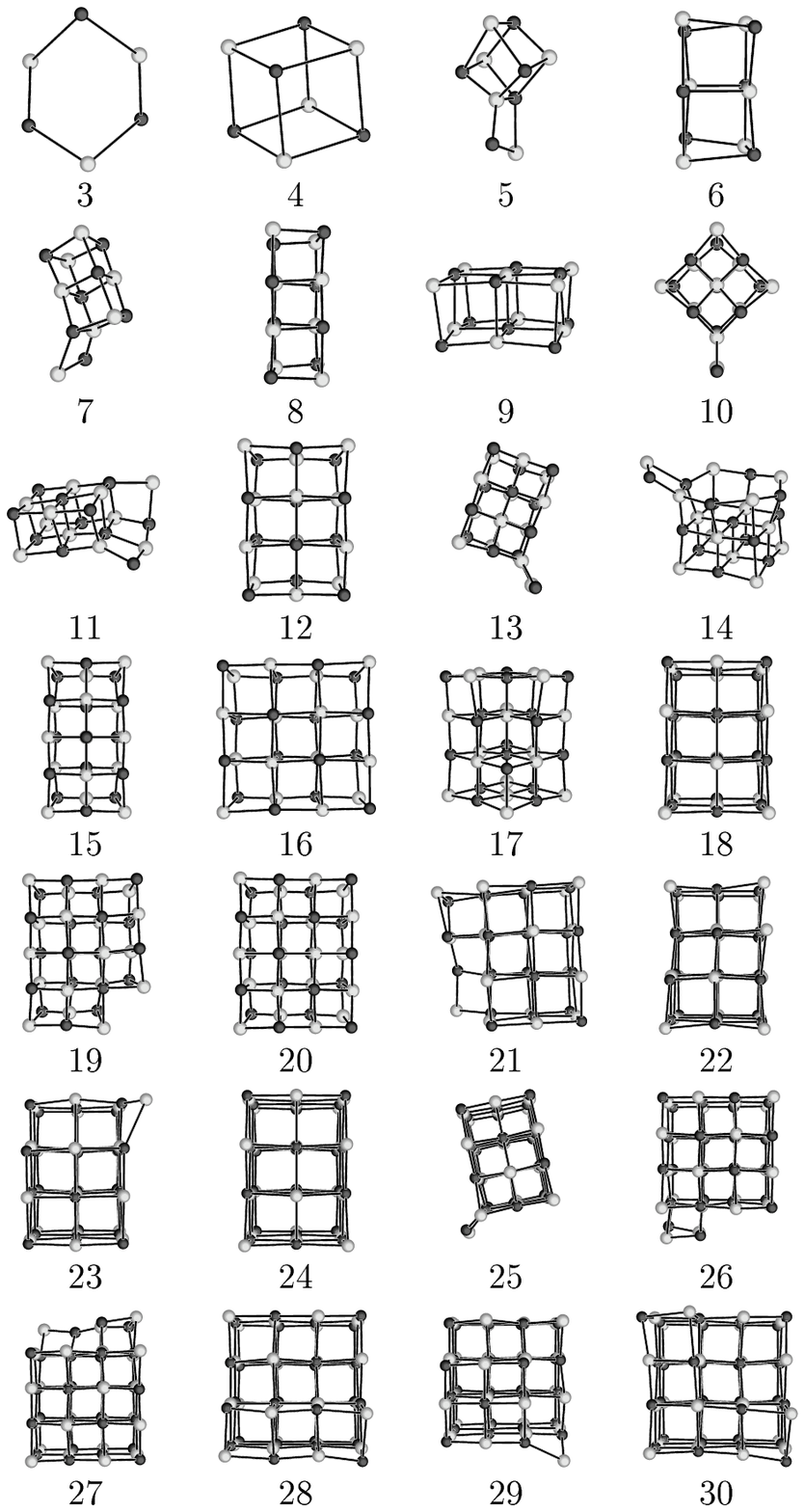}
\caption{Lowest energy structures of (MgO)$_n$ clusters from Monte Carlo
minimization using the self-consistent polarizable fluc-$q$ model.}
\label{fig:str}
\end{figure}

This set predicts the charge transfer to be 0.92 in the MgO molecule, and
1.92 in the crystal. The equilibrium distance is 1.86~\AA\ in the molecule
and the crystal lattice constant is 1.78~\AA. These data correspond to
respective errors of 6\%\ and 16\%\ when compared with the experimentally
measured values. The binding energies cannot be compared with reference
values, because the present model includes extra self-energy $U^0$ terms.

The lowest energy structures of (MgO)$_n$ clusters have been searched
using the basin-hopping or Monte Carlo minimization algorithm.\cite{lisch,bh}
For each size in the range $2\leq n\leq 30$, 5000 Monte Carlo steps were
performed starting from a random guess geometry. We also locally optimized
databases of structures found by global optimization of the rigid ion model
with fixed charges $\pm 1$ and $\pm 2$. In many cases, the global minimum was
found to lie within the database obtained with $Z=\pm 1$. 

The structures of the global minima are represented in Fig.~\ref{fig:str}.
Beyond $n=3$, they are based on small (MgO)$_4$ cubic units. The cuboid
picture found in the present Communication is essentially similar to the
results of Roberts and Johnston\cite{roberts} obtained with a genetic
algorithm, except for the slight distortions due here to polarization. We do
not find evidences for hollow\cite{ziemann,roberts} or ``spiky''
geometries.\cite{kohler} Stacking of (MgO)$_3$ hexagonal units leads to
isomers slightly less stable than cuboids at the same size. This partly
explains why the global minima at $n=14$ and $n=22$ differ from the results of
Roberts and Johnston.\cite{roberts} Actually the fact that hexagonal rings
are less favored in the present model is not in strong contradiction with
experimental results, for two reasons. Firstly, experiments have been
performed by mass spectrometry on charged species, which may well exhibit
different structures than neutrals.  Secondly, most
magic numbers peaks interpreted as the signature of hexagonal stacks are
indeed compatible with cuboid like geometries.

\begin{figure}[htb]
\setlength{\epsfxsize}{8.6cm}
\leavevmode \epsffile{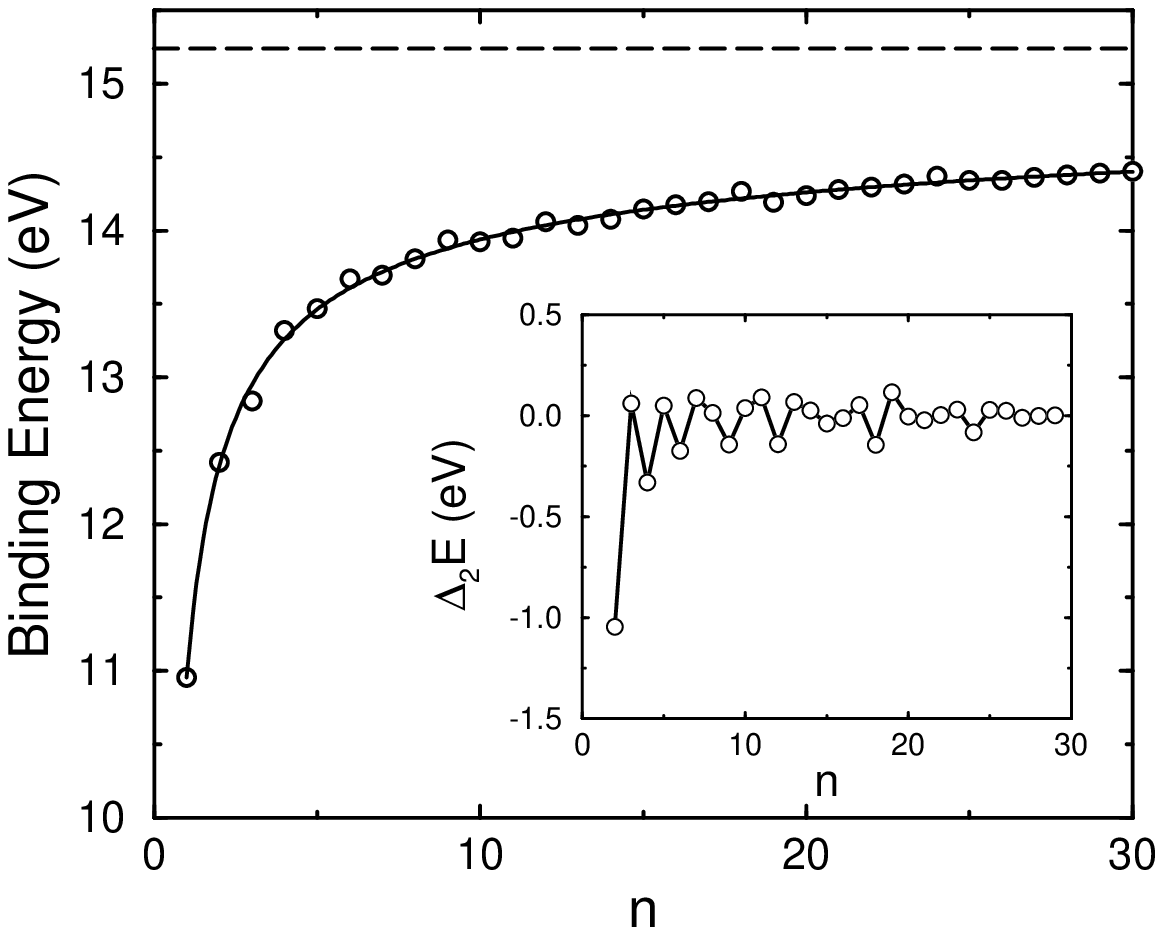}
\caption{Binding energy of (MgO)$_n$ clusters in the range $1\leq n\leq 30$.
The horizontal dashed line marks the asymptotic bulk limit. The open circles
correspond to the structures of Fig.~\protect\ref{fig:str}, the solid line
is a fit of the form $E(n)/n=a+bn^{-1/3}+cn^{-2/3}+dn^{-1}$, with parameters
given in the text. Inset: second energy difference $\Delta_2 E(n)=2E(n)-E(n+1)
-E(n-1)$ versus $n$.}
\label{fig:binding}
\end{figure}

The variations with size of the binding energy of the lowest energy structures
found with the present polarizable fluc-$q$ model are depicted in
Fig.~\ref{fig:binding}. The binding energy $E$ shows a global increase,
which can be fitted approximately in a liquid drop fashion as $E(n)
= -15.13n+1.83n^{2/3}+0.89n^{1/3}+1.45$. The values of the latter parameters
are slightly changed if we include larger cubic clusters such as (MgO)$_{108}$.
By construction, the crystal binding energy found from
this expression is close to the numerical minimization of Eq.~(\ref{eq:vbulk}).

As can be noted in Fig.~\ref{fig:binding}, there are some deviations from
the smooth behavior of the fitted energy. To see them more clearly, the second
energy difference $\Delta_2 E(n)=2E(n)-E(n+1)-E(n-1)$ has been represented in
the inset of this figure. This quantity is usually convenient to find the
special stabilities of some sizes. The most stable clusters appear here
at $n=2$, 4, 6, 9, 12, 18, and 24. The sizes $n=15$, 21, and 27 can be added
as relatively stable. All these clusters are perfect cuboids.
The magic character of the $n=15$ cluster is less marked, due to the fact
that (MgO)$_{16}$ is also a cuboid.

\begin{figure}[htb]
\setlength{\epsfxsize}{8.6cm}
\leavevmode \epsffile{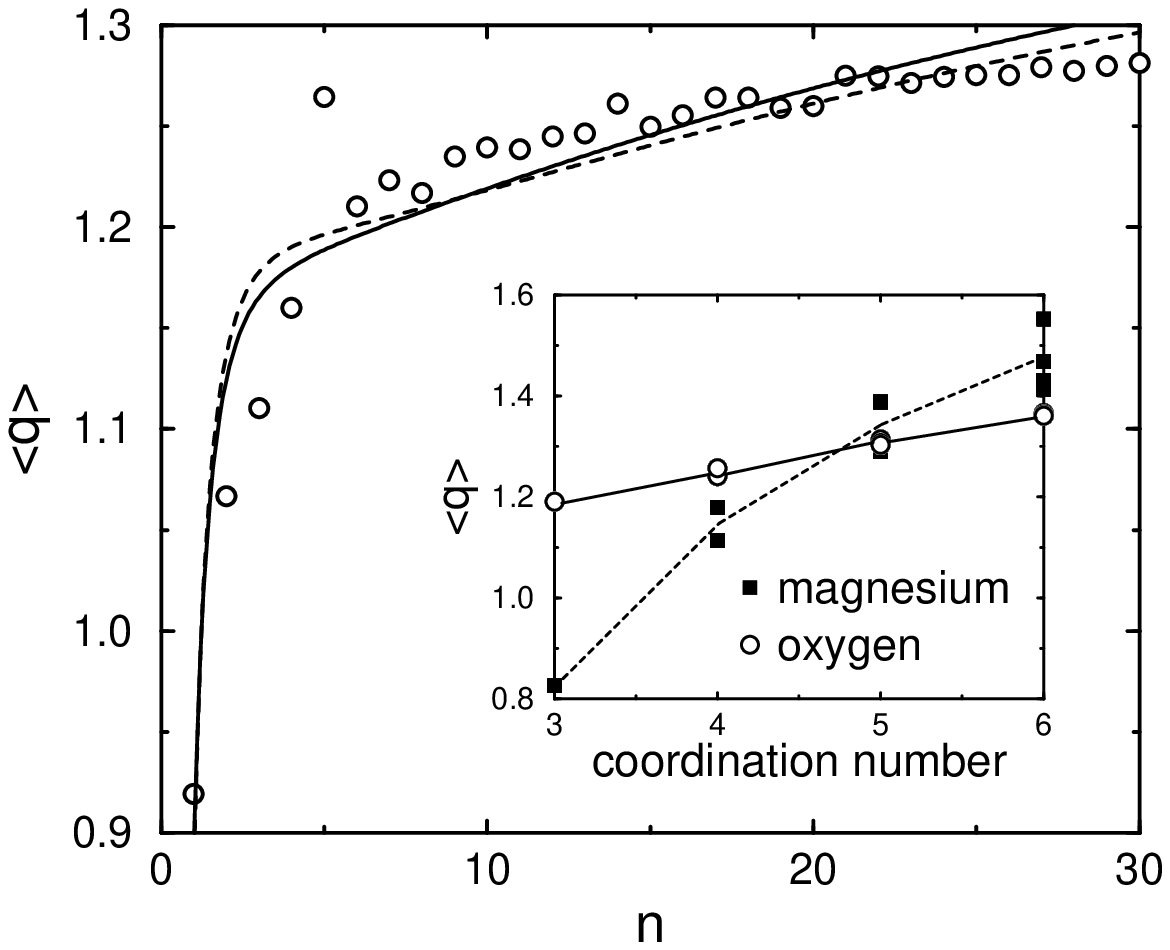}
\caption{Average modulus of the charge transferred over all ions in (MgO)$_n$
clusters. The lines are fits of the form $\langle q\rangle(n)=2-a'n^{-1/3}
-b'n^{-2/3}+c'n^{-1}$, with (solid line) or without (dashed line) the value
for the (MgO)$_{108}$ cubic cluster. Inset: effective moduli of the charges
carried by all magnesium (full squares) or oxygen (empty circles) ions in
the (MgO)$_{108}$ cluster versus their coordination number. The lines are a
guide to the eye.}
\label{fig:q}
\end{figure}

We turn now to the problem of charge transfer, and more generally to the
ionic or covalent nature of the chemical bonding in MgO clusters. In
Fig.~\ref{fig:q} we have represented the average charge $\langle q\rangle$
carried by the ions in the cluster, regardless of their position inside the
cluster or coordination number. This quantity is defined as the mean value
over all magnesium ions, which is the exact opposite of the mean value over
all oxygen ions. This
definition is somewhat loose and arbitrary, because all ions do not play the
same role in the cluster due to the large surface/volume ratio. From
Fig.~\ref{fig:q} we see that charge transfer is strongly size-dependent in the
present fluctuating charges model, and that the convergence toward the bulk
limit is much slower than assumed by K\" ohler and coworkers.\cite{kohler}
In fact, knowing that the average charge effectively reaches about 2 electrons
at large $n$ allows us to fit the variations of $\langle q\rangle$ with $n$
as $\langle q\rangle (n) \approx 2-a'n^{-1/3}-b'n^{-2/3}-c'n^{-1}$. The
effective crossover size between mixed ionic/covalent bonding and pure ionic
bonding can then be estimated as $n^*$ such that $\langle q\rangle (n^*)\sim
3/2$.
To get more realistic values for $n^*$, we have included the data
corresponding to larger cubic clusters, namely (MgO)$_{32}$ and especially the
$6\times 6\times 6$ cluster (MgO)$_{108}$. Because the size range covered
remains relatively small, the value for the latter cluster was given a weight
of 10 in the fitting process, with respect to smaller sizes. Using this
procedure, we find the crossover size to be located at $n^*\sim 300\pm 100$
depending on the presence of the large cluster in the fit. Including the
value for (MgO)$_{108}$ results in an increase of $n^*$, and including
the data for larger clusters should further shift the crossover
size toward several hundreds or thousands MgO molecules.

Within the present empirical model, electrostatic properties are naturally
coordination dependent. In the inset of Fig.~\ref{fig:q}, the modulus of the
charge transferred is plotted for each ion in the (MgO)$_{108}$ nanocrystal
versus its coordination number. Several features are of interest. First, 
the magnitude of the charge transferred increases with coordination, as
expected from the decreasing intensity of the electric field. This is in
agreement with the electronic structure calculations performed by Recio {\em
et al.},\cite{recio} by Veliah {\em et al.},\cite{veliah} and more recently
by Coudray {\em et al.}\cite{coudray} Second, the charge carried by magnesium
ions is more sensitive to coordination than the charge carried by oxygen ions.
This is also in agreement with the findings of Veliah {\em et al.}\cite{veliah}
The above results confirm that conventional potentials with fixed
charges are not fully appropriate to describe MgO clusters. This had been
addressed by Wilson who considered
phenomenological coordination-dependent polarizabilities within the
compressible-ion model.\cite{wilson} Coordination-dependent charges are a
natural outcome of the present potential, allowing to study MgO clusters in a
wide range of condensed phases. While the present potential is able
to treat large clusters beyond the possibilities of first principles-based
computations, the needed inversion of a square matrix can be a limiting
factor. Fortunately, extended Lagrangian techniques\cite{rick} can reduce the
computational cost significantly, making the polarizable fluc-$q$ model
valuable in thermodynamical context.

To conclude, we proposed an empirical model to describe ionic/covalent
bonding in MgO clusters. This model is based on fluctuating charges and
incorporates atomic polarization in a self-consistent way. By fitting it
on both molecular and bulk properties we found that small clusters
preferentially exhibit cuboid geometries, showing magic numbers
in good agreement with experiments. The average charge carried by magnesium
or oxygen atoms smoothly increases, and the crossover between ionic/covalent
and pure ionic bonding was estimated to be above 300 molecules. The model
correctly predicts that the charge transferred depends on coordination.

\end{document}